# Precursor Ratio-Controlled Chloride Vacancies in Lead-Free $KCuCl_{3-x}$ Memristors for Forming-Free Gradual Resistive Switching and Neuromorphic Learning

Chinmayee Mandar Mhaskar[1], Sangeeta Roy Chaudhuri[2,*], Ayan Roy Chaudhuri[1,*]

[1] Materials Science Centre, Indian Institute of Technology Kharagpur, India 721302.

[2] Department of Chemistry, Raja Narendralal Khan Women's College, Midnapore, India 721102.

[*]Email ID: ayan@matsc.iitkgp.ac.in, src_chemistry@rnlkwc.ac.in

## Abstract

Defect engineering of memristive materials offers a powerful strategy for regulating ionic transport and analog resistive switching required for neuromorphic computing. Here, we report a wet chemical synthesis route for chloride-vacancy ($V_{Cl}$) engineering in lead-free all-inorganic $KCuCl_{3-x}$ thin films and establish a direct correlation between defect concentration, electronic structure, transport dynamics, and neuromorphic performance. Controlled precursor-ratio variation systematically tunes the $V_{Cl}$ concentration in $KCuCl_{3-x}$ thin films, driving preferential vacancy formation at the terminal chloride sites, band-gap narrowing, and enhanced electronic disorder. W/$KCuCl_{3-x}$/FTO memristors exhibit stable bipolar VCM-type resistive switching, with increasing $V_{Cl}$ concentration promoting forming-free operation. An optimal $V_{Cl}$ concentration (~ $1.3\times10^{21}$ $cm^{-3}$) provides a balance between ionic mobility and defect controllability, resulting in highly linear potentiation/depression characteristics and superior pulse endurance ($>10^4$ cycles). When the experimentally measured conductance characteristics of the optimized memristor are implemented as synaptic weights in artificial neural network simulations, recognition accuracies of 98.96% and 91.82% are achieved on the MNIST and Fashion-MNIST datasets, respectively. These findings establish precursor ratio-controlled $V_{Cl}$ engineering as an effective strategy for simultaneously optimizing ionic transport, analog switching, and neuromorphic learning in lead-free halide perovskite memristors, providing a general materials design principle for energy-efficient neuromorphic electronics.

## 1. Introduction

Point defects play a fundamental role in governing the electronic structure, redox equilibria, and ionic transport properties of functional materials. In mixed ionic-electronic conductors, the formation energy, charge state, and migration barrier of intrinsic vacancies determine carrier density, ionic mobility, and electrochemical response.[1] Under an applied electric field, charged vacancies in ionic solids can be driven far from equilibrium, enabling defect redistribution and coupled redox reactions that underpin intrinsic resistive switching (RS) through a valence-change mechanism (VCM). Consequently, precise control over defect concentration and migration dynamics is essential not only for tuning conventional RS characteristics, including switching voltage, resistance window, endurance and retention, but also for achieving stable analog conductance modulation required to emulate biological synaptic plasticity.

The performance of VCM-based memristive electronic synapses is intrinsically linked to defect-mediated transport and redistribution. Highly linear and symmetric potentiation/depression (P-D) characteristics are essential for accurate synaptic weight updates during neural network training, whereas nonlinear conductance evolution leads to cumulative programming errors and reduced recognition accuracy.[2]

The kinetics of ionic defects further determine the programming voltage, pulse duration, and energy required for conductance modulation.[3] While insufficient defect concentrations in the RS materials limit ionic transport, resulting in slow switching and high programming energy, excessive defect concentrations promote vacancy aggregation, uncontrolled filament evolution, conductance saturation, poor switching reversibility, and degraded P-D linearity.[2] These competing effects highlight the need for deterministic defect engineering in memristive materials to establish an optimal defect landscape for neuromorphic applications.

Halide perovskites (HPs) have emerged as an attractive materials platform for defect-mediated RS owing to their soft ionic lattices, low halide-vacancy formation energies, and highly tunable defect chemistry, which facilitate efficient ionic migration under moderate electric fields.[4–6] Additionally, their low-temperature solution-processable synthesis offers a scalable and cost-effective route for device fabrication.[7–10] Despite these advantages, establishing quantitative relationships between defect chemistry, ionic transport in the HPs, and their neuromorphic functionalities remains a major challenge.[11–13]

Most reported HP memristors employ environmentally hazardous Pb-containing compositions or hybrid organic-inorganic perovskites. Furthermore, many devices rely on electrochemically active metal electrodes, in which resistive switching is dominated by electrochemical metallization rather than by intrinsic halide-vacancy migration.[14–16] As a result, the intrinsic role of halide vacancies in governing the evolution of analog conductance remains insufficiently understood.[12,17–21]

Lead-free all-inorganic halide perovskites (LFAIHPs), which exhibit intrinsic vacancy-mediated VCM, therefore represent an ideal model system for elucidating the fundamental role of ionic defects in determining the RS properties of environmentally benign memristive materials.[22] Among them, copper-based halide perovskites are particularly attractive, since halide vacancy formation is intrinsically coupled to the $Cu^{2+}/Cu^{+}$ redox equilibrium, [19,21,23] providing a direct link between defect chemistry and electronic transport.

We recently demonstrated stable analog RS driven by a chloride-ion vacancy ($V_{Cl}$) in a single-composition $KCuCl_{3-x}$ memristor, establishing this LFAIHP material as a promising intrinsic VCM platform.[23] However, the dependence of ionic transport kinetics, analog conductance evolution, and, ultimately, the modulation of neuromorphic learning performance of $KCuCl_{3-x}$ memristors on the concentration of $V_{Cl}$ remains to be elucidated. More broadly, quantitative correlations linking vacancy concentration, defect-mediated transport, synaptic characteristics, and artificial neural network (ANN) performance remain largely unexplored in HP-based memristive systems.

Here, we establish a precursor-ratio-controlled vacancy-engineering strategy to systematically tune the $V_{Cl}$ concentration in $KCuCl_{3-x}$ thin films and correlate defect chemistry with transport and neuromorphic functionality. Structural refinement combined with X-ray photoelectron spectroscopy supports preferential vacancy formation at terminal chloride sites across five precursor-controlled compositions. Temperature-dependent electrical transport, impedance spectroscopy, and pulse-programming measurements reveal how chloride vacancies regulate ionic migration and analog conductance modulation. Finally, by directly correlating vacancy concentration with synaptic characteristics and ANN recognition accuracy, we establish a quantitative defect-transport-learning relationship that identifies precursor-controlled vacancy engineering as an effective route for developing high-performance, lead-free halide perovskite memristors for neuromorphic computing. Hence, unlike previous studies that only demonstrate resistive switching in halide perovskites, this work establishes a correlation between defect transport and neuromorphic functionality through systematic $V_{Cl}$ engineering.

## 2. Results and Discussion

Precursor stoichiometry was consistently varied to evaluate its influence on the structural evolution and stoichiometry of $KCuCl_{3-x}$ thin films. The KCl: $CuCl_2$ molar ratio was varied from 1:1 to 0.6:1 (**Table 1**), while all deposition and annealing parameters were kept constant.

**Table 1.** Variation in precursor ratio for $KCuCl_{3-x}$ films synthesis

| Sample Name | KCl:$CuCl_2$ Molar ratio |
|---|---|
| HP1 | 1:1 |
| HP2 | 0.9:1 |
| HP3 | 0.8:1 |
| HP4 | 0.7:1 |
| HP5 | 0.6:1 |

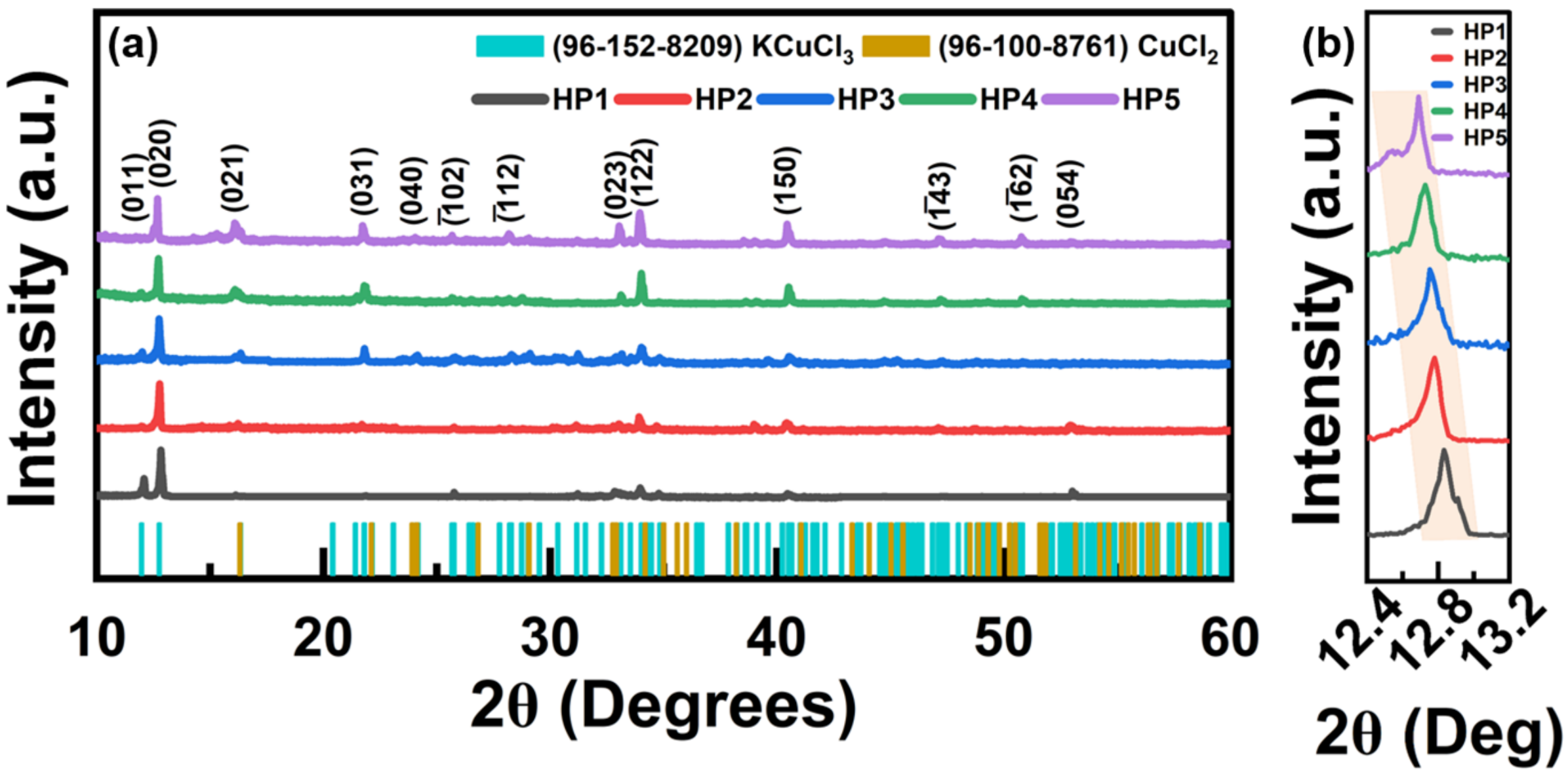

**Figure 1. (a)** XRD patterns of $KCuCl_{3-x}$ thin films compared with $KCuCl_3$ and $CuCl_2$ PDF cards from the database, and **(b)** Enlarged view of the perovskite phase XRD peak for the (020) plane, across HP1 to HP5 samples.

**Figure 1a** illustrates the XRD patterns of the $KCuCl_{3-x}$ thin films in ω-2θ geometry, together with reference diffraction patterns of $KCuCl_3$ (JCPDS card no. 96-152-8209) and $CuCl_2$ (JCPDS card no. 96-100-8761).

All compositions exhibit diffraction peaks which can be indexed to monoclinic $KCuCl_3$, confirming retention of the crystal structure across the compositional series (**Figure S1**). The diffraction peaks shift systematically towards lower 2θ angles from HP1 to HP5, as shown in **Figure 1b** for the most intense peak, corresponding to the (020) plane at 2θ ≈ 12-13°. Such a shift of the diffraction peaks indicates progressive lattice expansion with a reduction in the KCl precursor ratio.

Rietveld refinement was performed to quantify structural changes across the HP1-HP5 series. The refined lattice parameters, atomic positions, and occupancies are summarized in **Section 2 of the SI**. All compositions crystallize in the monoclinic structure (space group $P2_1/c$) comprising edge-sharing $Cu_2Cl_6$ units that form a rigid structural framework (**Figure S2**).[24] The *a* and *c* axes decrease slightly from 4.028 ± 0.0002 Å to 4.017 ± 0.0002 Å and 8.734 ± 0.0004 Å to 8.712 ± 0.0004 Å, for HP1 and HP5, respectively. In contrast, the *b* axis expands substantially from 13.789 ± 0.0005 Å to 13.946 ± 0.0005 Å. Concomitantly, the monoclinic angle (β) decreases systematically from 97.15 ± 0.01° (HP1) to 96.68 ± 0.01° (HP5), resulting in an increase in the unit cell volume, V = 481.33 $Å^3$ (HP1) to 484.72 $Å^3$ (HP5) (Table S1). The anisotropic lattice relaxation is dominated by the ~1.14% expansion along the *b* axis, which outweighs the minor contractions along the *a* and *c* axes, which is reflected in the systematic shift of the (020) peak toward lower 2θ values (Figure 1b).

To elucidate the origin of the structural distortion, crystallographic site occupancies were examined by Rietveld refinement of the X-ray diffraction data. All cation sites (K and Cu) refined to full occupancy within experimental uncertainty (±0.02), with no evidence of cation vacancies, antisite disorder, or interstitial incorporation. In contrast, the occupancy refinement indicates a systematic reduction at the terminal chlorine site ($Cl_T$, labelled Cl(3) (**Figure S2**) with decreasing KCl precursor concentration. Specifically, the refined occupancy of Cl(3) decreases monotonically from 0.87 ± 0.02 (HP1) to 0.77 ± 0.02 (HP5), corresponding to average compositions of $KCuCl_{3-x}$, where x varies from 0.13 to 0.23, based on refined site occupancies. The observed change in occupancy substantially exceeds the associated uncertainties, indicating that the trend is statistically significant and intrinsic to the crystal structure rather than an artefact of the refinement. The preferential formation of vacancies at the $Cl_T$ is chemically consistent with the weaker semi-coordinate Cu-$Cl_T$ bonding relative to the bridging Cu-$Cl_B$-Cu interactions that stabilize the $[Cu_2Cl_6]^{2-}$ framework. [24–26]

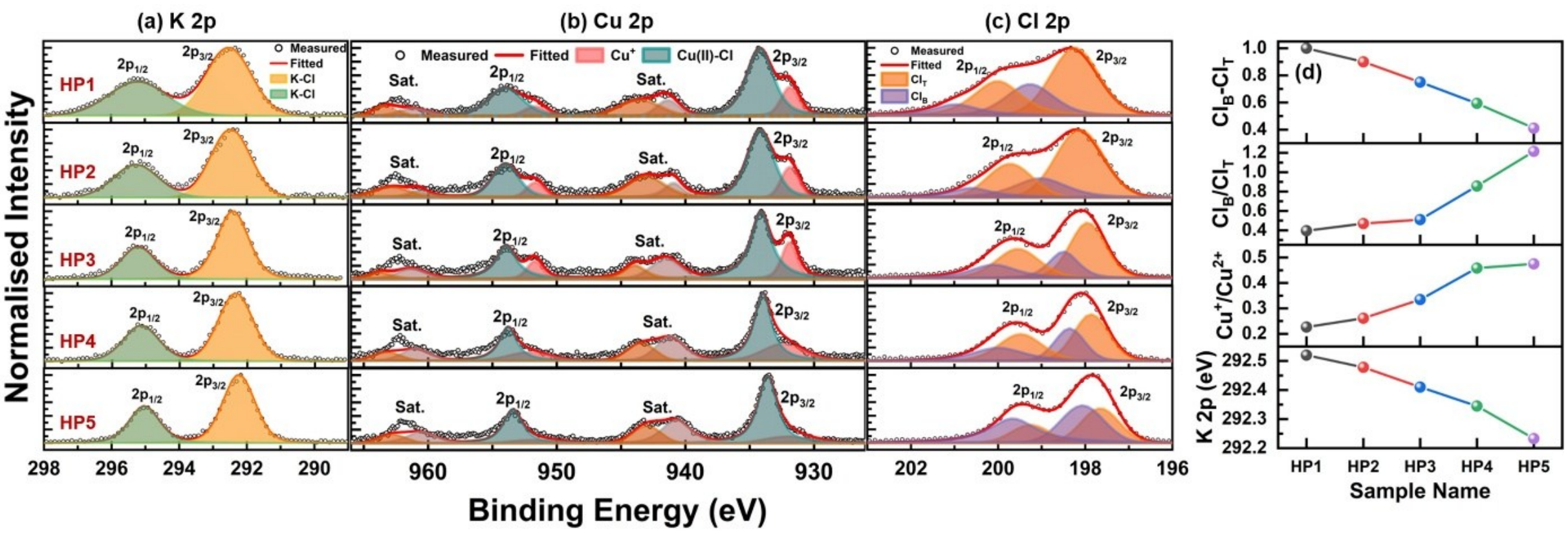

**Figure 2.** High-resolution XPS elemental scan for **(a)** K 2p, **(b)** Cu 2p, **(c)** Cl 2p regions across HP1-HP5 compositional series and **(d)** variations in K2p peak shift, $Cu^+/Cu^{2+}$, $Cl_B/Cl_T$ peak area ratios, difference in $Cl_B$ and $Cl_T$ binding energies across HP1-HP5.

**Figure 2 (a-c)** presents the high-resolution XPS spectra of the K 2p, Cu 2p, and Cl 2p regions for all five compositions. All binding energies were calibrated with reference to the C 1*s* peak at 284.5 eV, which corresponds to adventitious carbon. The evolution of K 2p binding energy, $Cu^+/Cu^{2+}$ area ratio, $Cl_B/Cl_T$ area ratio, and the binding energy difference between $Cl_B$ and $Cl_T$ is summarized in **Figure 2d.**

The K 2p spectra exhibit a characteristic spin-orbit splitting of ~2.8 eV between the $2p_{3/2}$ and $2p_{1/2}$ peaks.[27] With increasing $V_{Cl}$ concentration, the K $2p_{3/2}$ peak shifts from 292.5 eV (HP1) to 292.23 eV (HP5), consistent with a reduction in the local Madelung potential and enhanced electronic screening resulting from $V_{Cl}$-induced charge redistribution.[28]

The Cu 2p spectra were deconvoluted into $Cu^{+}$, $Cu^{2+}$ and $Cu^{2+}$ shake-up satellite components.[23] HP1 exhibits prominent $Cu^{2+}$ satellite features at 941-944 eV and 961-963 eV, characteristic of the $3d^9$ electronic configuration of $Cu^{2+}$. Across the HP1-HP5 series, the $Cu^{2+}$ satellite intensity progressively decreases, accompanied by a monotonic increase in the $Cu^{+}/Cu^{2+}$ peak-area ratio, indicating gradual electron redistribution within the Cu-Cl framework as the $V_{Cl}$ concentration increases.[29,30]

The Cl 2p spectra were deconvoluted into $Cl_T$ and $Cl_B$ components.[31] For HP1, the $Cl_T$ and $Cl_B$ $2p_{3/2}$ peaks are located ~198.02 and ~199.02 eV, respectively, with a binding energy separation of ~1eV. The higher binding energy of $Cl_B$ originates from its bridging Cu-$Cl_B$-Cu coordination, where interaction with two neighbouring Cu centres generates a stronger local electrostatic field than that experienced by $Cl_T$.[32] With increasing $V_{Cl}$ concentration, peaks corresponding to both chloride species shift toward lower binding energies; although, the magnitude of the shift is substantially larger for $Cl_B$ (~0.95 eV) than for $Cl_T$ (~0.36 eV). Consequently, the binding energy separation between $Cl_B$ and $Cl_T$ decreases from ~1 eV in HP1 to ~0.41 eV in HP5. This pronounced reduction in the $Cl_B$-$Cl_T$ energy difference is consistent with stronger vacancy-induced electronic redistribution within the Cu-$Cl_B$-Cu framework, resulting in significant modification of the local electrostatic environment around the bridging chloride ions.[32,33]

Furthermore, a progressive increase in the $Cl_B/Cl_T$ peak area ratio with increasing vacancy concentration is observed, which suggests a preferential vacancy at the $Cl_T$ site.

Quantitative analysis of the XPS atomic percentages (**Table S5**) yields compositions ranging from $KCuCl_{2.89}$ (HP1) to $KCuCl_{2.79}$ (HP5), in excellent agreement with Rietveld-refined compositions, thereby providing independent validation of the $V_{Cl}$ concentrations (**Table S6**).

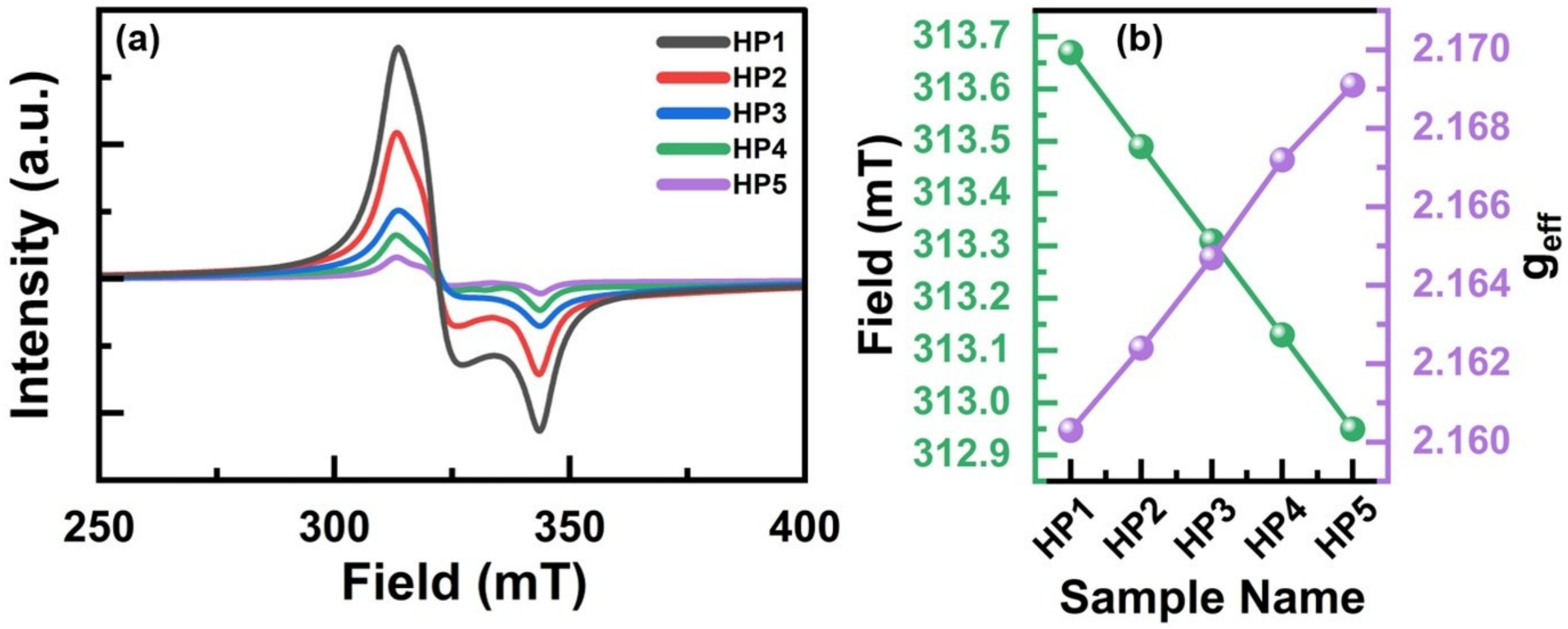

**Figure 3.** EPR spectra of the prepared samples showing variations in **(a)** signal intensity and **(b)** effective g-factor (Reference lines have been added to aid the visualization).

The XPS analysis suggests that increasing $V_{Cl}$ concentration is accompanied by progressive reduction of $Cu^{2+}$ centres. To independently verify this vacancy-induced evolution of Cu oxidation state in the bulk of the $KCuCl_{3-x}$ samples, EPR spectroscopy was performed on all five compositions using the same amount of powder (20 mg). Details of the measurements have been described in **section S5** of the SI.

All samples exhibit the characteristic EPR signal of $Cu^{2+}$; however, the first derivative EPR spectra reveal a systematic decrease in signal intensity from HP1 to HP5 (**Figure 3a**). Since the EPR signal intensity is directly proportional to the concentration of paramagnetic centres, under identical experimental conditions, the progressive attenuation of the $Cu^{2+}$ signal reflects a corresponding decrease in the $Cu^{2+}$ population as the KCl precursor ratio decreases. This observation, along with the crystallographic refinement and XPS studies, establishes that charge compensation with increasing $V_{Cl}$ concentration in $KCuCl_{3-x}$ occurs through electron localization at Cu, reducing paramagnetic $Cu^{2+}$ ($3d^9$) to EPR-silent $Cu^+$ ($3d^{10}$).

In addition to the reduction in signal intensity, the resonance field shifts monotonically from 313.67 mT for HP1 to 312.95 mT for HP5 (**Figure 3b and S3**), corresponding to an increase in the effective g-factor from 2.1603 to 2.1691 (**Figure 3b and Table S7**). The observed variation in g ($\Delta g \approx 0.009$) is significantly higher than the instrumental field uncertainty (±0.1-0.2 mT, corresponding to $\Delta g \approx 0.0008$), excluding measurement artefacts and confirming that the shift is intrinsic to the material.[34] Given that the g-factor is highly sensitive to the local crystal field and spin-orbit coupling surrounding $Cu^{2+}$ centres, its systematic increase reflects progressive modification of the electronic environment arising from vacancy-induced electron redistribution.[35]

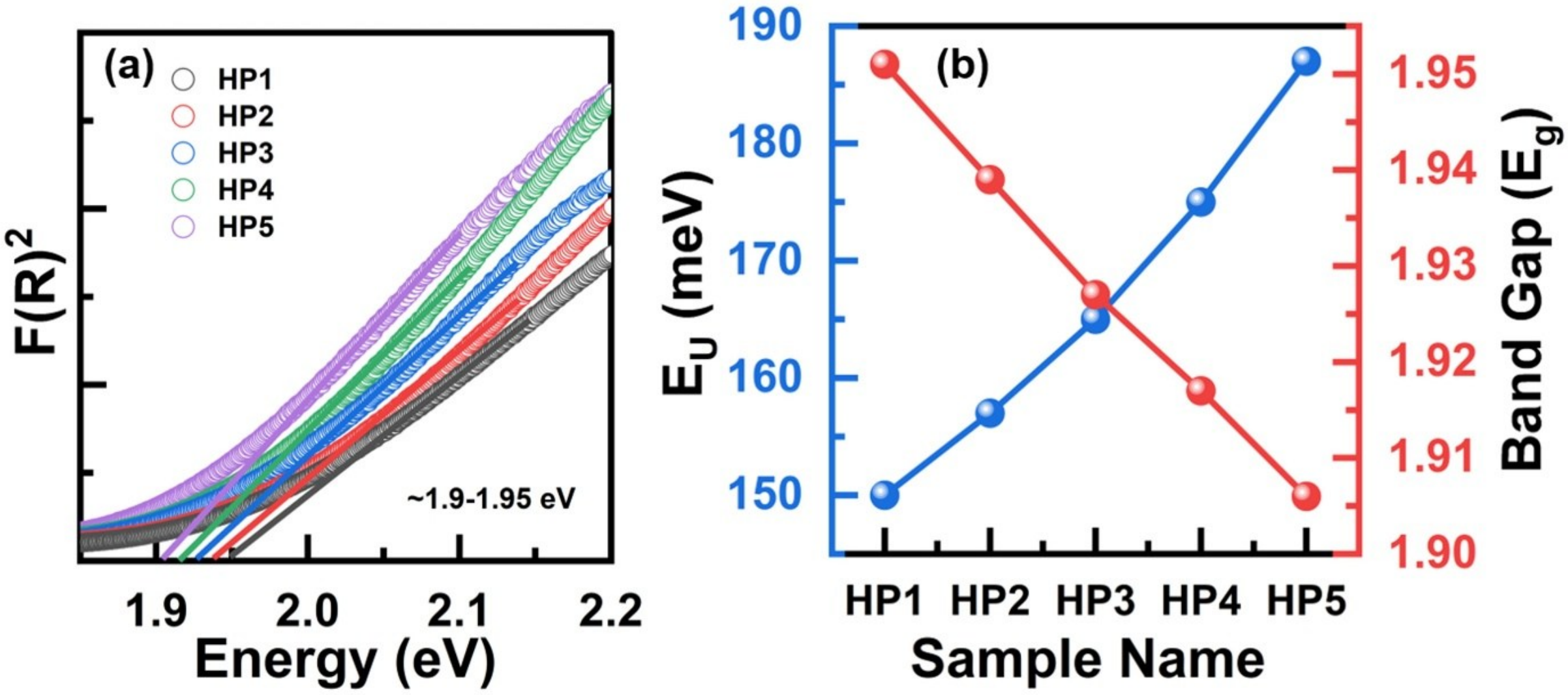

**Figure 4. (a)** Tauc plots depicting extrapolation of absorption data to determine the bandgap of the prepared samples, and **(b)** Urbach energy ($E_U$) values derived from the tail of UV-Vis spectra and band gap $E_g$ trend across HP1-HP5 compositional series (Reference lines have been added to aid the visualization).

To elucidate the influence of $V_{Cl}$ on the electronic structure, UV-Vis-DRS spectroscopy was performed across the HP1-HP5 compositional series. The reflectance data were transformed using the Kubelka-Munk function [36,37]:

$$F(R) = \alpha = \frac{(1-R)^2}{2R} \quad (1)$$

where $\alpha$ is the absorption coefficient, and $R$ is the percentage of reflected light. The optical band gap energy ($E_g$) and incident photon energy ($h\nu$) are related according to the transformed Kubelka-Munk function[36,37],

$$F(R)h\upsilon^{1/\rho} = A(h\upsilon - E_g) \quad (2)$$

where $A$ is the transition probability-dependent constant and $\rho$ is an index associated with the nature of the band energy transition.[27] The value of ρ equals 2 or $^1/_2$ for an indirect or a direct allowed transition, respectively. HP1 exhibits a direct band gap of 1.95 eV (**Figure 4a)**, consistent with previously reported values for $KCuCl_3$-based systems.[37]

A systematic narrowing of the band gap from ~1.95 eV (HP1) to ~1.9 eV (HP5) is observed (**Figure 4b**). This monotonic trend closely parallels the reduction in $Cl_T$ occupancy and the progressive $Cu^{2+}$→$Cu^{+}$ conversion, demonstrating that the electronic structure evolves systematically with increasing $V_{Cl}$ concentration. In copper chloride-based systems, the upper valence band is predominantly constituted by the hybridization of Cl 3p orbitals and Cu 3d states.[38] Consequently, the formation of $V_{Cl}$ perturbs the local Cu-Cl bonding environment, introducing defect states near the valence-band edge and shifting the band edges, thereby producing the observed band-gap narrowing. Simultaneously, the vacancy-induced partial reduction of $Cu^{2+}$ to $Cu^{+}$ modifies the Cu 3d electronic configuration and redistributes the density of states near the band edges.[37,39] These coupled structural and electronic modifications demonstrate that $V_{Cl}$ act as electronically active defects that govern the evolution of the band structure.[40]

Additional insight into vacancy-induced electronic disorder was obtained from analysis of the Urbach energy tails ($E_U$) using:

$$\propto = \propto_0 \exp\left(\frac{h\nu - E_g}{E_U}\right) \quad (3)$$

Where α is the absorption coefficient, h is Planck's constant, ν is the frequency, Eg is the band gap, and $E_U$ is the Urbach energy.[41] The extracted $E_U$ increases progressively from HP1 to HP5 (**Figure 4b**), indicating an increasing density of disorder-induced electronic states with increasing $V_{Cl}$ concentration. [42]

The combined structural and spectroscopic analyses demonstrate that precursor-ratio-controlled $V_{Cl}$ engineering modifies both the crystal and electronic structures of $KCuCl_{3-x}$ thin films.

Atomic force microscopy (AFM) and cross-sectional scanning electron microscopy (SEM) investigations were carried out to elucidate the morphological features of the HP1-HP5 series. All the films exhibit dense, uniform surfaces with a low root-mean-square (RMS) roughness of ~5 nm (**Figure S4**) and a nearly uniform thickness of ~350 nm (**Figure S5**). Kelvin probe force microscopy (KPFM) was performed under identical experimental conditions for all compositions to compare their surface-potential landscape prior to investigating the electrical transport characteristics.

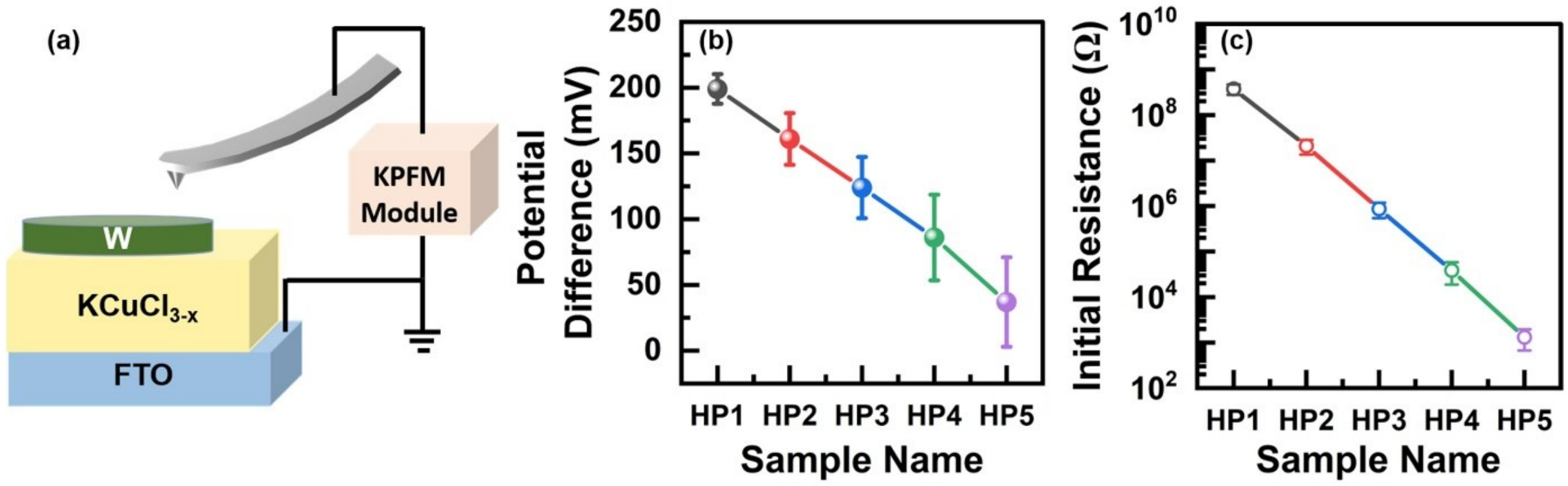

**Figure 5. (a)** Schematic representation of the KPFM measurement setup, **(b)** Potential difference estimated from KPFM CPD maps across the W/HP1-W/HP5 interfaces represented in Figure S6f-j, and **(c)** Pristine-state resistance of W/$KCuCl_{3-x}$/FTO devices measured at +0.1 V before electroforming or resistive switching. Error bars represent the standard deviation, and lines are drawn to aid the visualization.

**Figure 5a** represents the schematic of KPFM setup employed to scan the W/HP1-HP5 surfaces. **Figure 5b** plots the relative surface potential difference between the W electrode compared to the $KCuCl_{3-x}$ films, which decreases systematically from HP1 to HP5. Since the contact potential difference (CPD) obtained from the KPFM measurements reflects the contact potential difference associated with Fermi level alignment between the probe-tip and sample,[43,44] the observed trend suggests a systematic decrease in $KCuCl_{3-x}$ work function with increasing concentration of the $V_{Cl}$s. Complementary analysis of the valence band edge from XPS measurements (**Figure S6 a-e**) shows that the Fermi level ($E_F$) progressively shifts toward the conduction-band edge from HP1 to HP5, indicating increasing n-type character with increasing $V_{Cl}$ concentration (**Table S8**). This evolution agrees with the donor-type character commonly associated with halide ion vacancies in perovskites [24].

The increase in $V_{Cl}$ concentration is accompanied by a monotonic decrease in the initial resistance of devices from HP1 to HP5 (**Figure 5c**), consistent with the increase in carrier concentration associated with the progressively increasing doping of the films. [44–46]

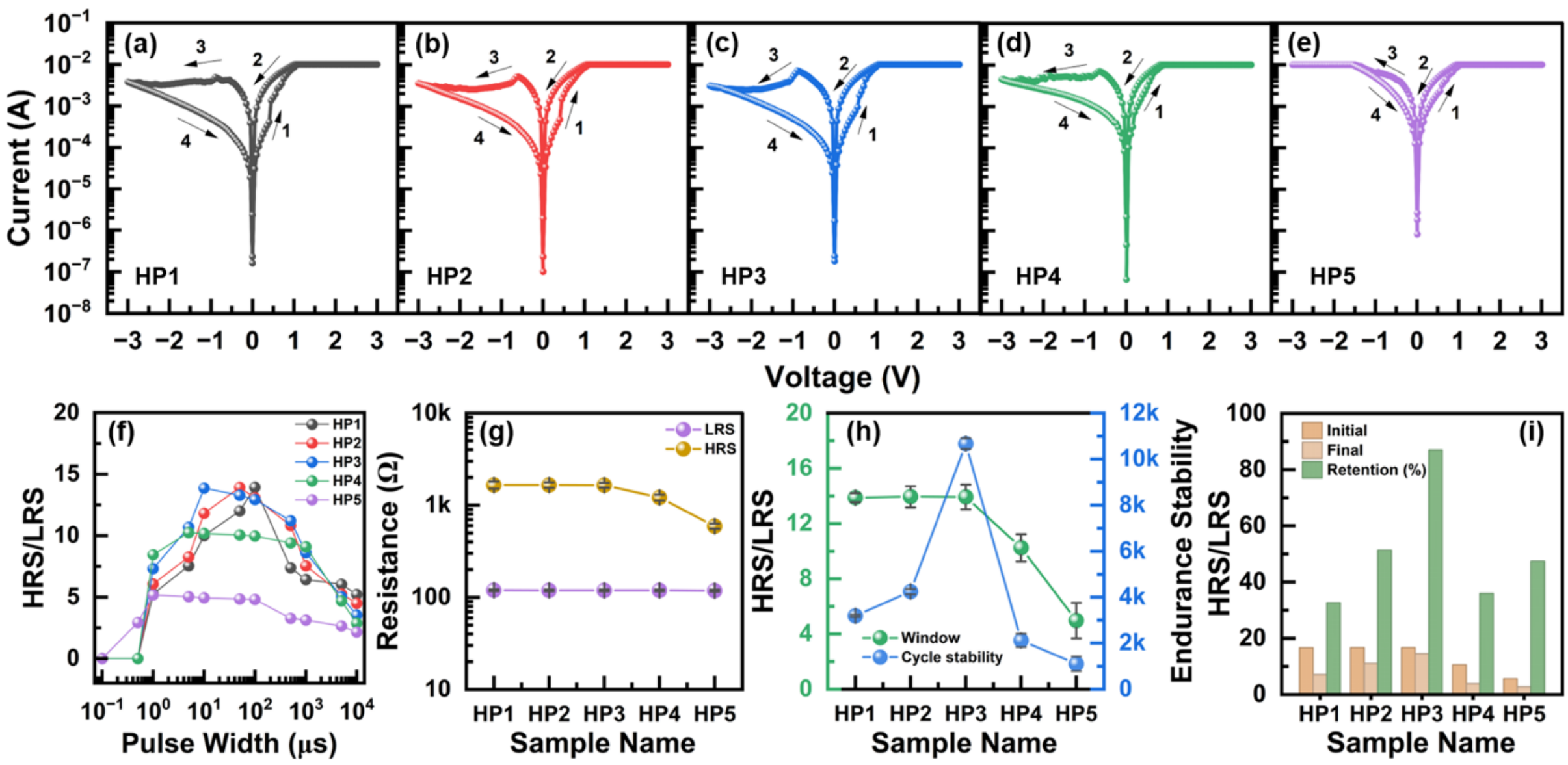

**Figure 6. (a-e)** DC I-V characteristics (median cycle of 100 repetitive cycles is represented and cycle-to-cycle variation is depicted in **Figure S7**), **(f)** pulse width dependent HRS/LRS variation, **(g)** device-to-device variability of LRS and HRS stability in pulse endurance study (error bars are shown to represent the standard deviation), **(h)** HRS/LRS ratio and endurance cycles stability, and **(i)** initial and final HRS/LRS as well as corresponding retention % after $10^4$ s, across memristors fabricated on HP1-HP5 samples.

The influence of $V_{Cl}$ concentration on resistive switching was first investigated by DC current-voltage (I-V) measurements using W/$KCuCl_{3-x}$/FTO devices. The voltage was swept in the sequence 0.0 V → +3V → 0.0 V → -3V → 0.0 V, with a step size of 0.05 V and a compliance current of 10 mA. All devices exhibited reproducible, bipolar, non-volatile RS and stable transitions between clearly separated high-resistance state (HRS) and low-resistance state (LRS) over multiple cycles (**Figure 6 and S7).** Notably, HP1 and HP2 require an initial electroforming step (+6 V and +4 V, respectively), whereas HP3-HP5 exhibit forming-free resistive switching. The elimination of the electroforming process indicates that increasing the $V_{Cl}$ concentration progressively facilitates the formation of conductive pathways in the $KCuCl_{3-x}$ memristors.

The nature of the set process also changes systematically across the compositional series. HP1 and HP2 devices exhibit an abrupt increase in current at ~ +0.5 V, consistent with the rapid formation of localised conductive filaments.[23,47] In contrast, HP3-HP5 exhibit a gradual increase in current, indicative of analog RS with progressive conductance evolution. This transition from abrupt to gradual switching demonstrates that precursor-ratio-controlled vacancy engineering enables controlled modulation of the switching dynamics while preserving the intrinsic bipolar valence-change switching mechanism previously established for $KCuCl_{3-x}$ memristors.[23]

The resistance states of the $KCuCl_{3-x}$ devices systematically evolve with increasing $V_{Cl}$ concentration. While the LRS remains nearly invariant across the HP1-HP5 series, the HRS decreases progressively from HP1 to HP5, reducing the resistance window from approximately 14 to 5. The decrease in HRS resistance is attributed to the enhanced bulk conductivity resulting from the increasing vacancy concentration, whereas the conductivity of the vacancy-rich filaments driving the LRS current remains largely unchanged.

Electrode area-dependent resistance measurements were performed across the HP1-HP5 series to elucidate the nature of the conductive pathways (**Figure S8**). The resistance scaling was analyzed using an empirical power law ($R=CA^{m-1}$), which provides insight into the spatial distribution of the current flow, specifically distinguishing between localised filamentary transport (m=1) and area-distributed homogeneous conduction (m=0).[48–50] Our analysis indicates predominantly filamentary conduction for all the devices. For the LRS, $m_{LRS}$ remains close to unity for HP1-HP3 and decreases progressively with increasing $V_{Cl}$ concentration (**Table S9**), indicating an increasing contribution of area-dependent conduction at higher vacancy concentrations. This trend is consistent with the switchover from highly localised filament formation to the formation of multiple filaments ($0<m<1$) with increasing $V_{Cl}$ concentration. [50] The HRS exhibits a progressively increasing scaling factor (m) from HP1 to HP5, showing a convergence trend with LRS, further suggesting that large $V_{Cl}$ concentration promotes a more spatially distributed conductive landscape rather than a highly localised switching pathway.

Temperature-dependent resistance measurements (**Figures S9 and S10**) across HP1-HP5 exhibit a negative temperature coefficient of resistance at HRS, characteristic of thermally activated semiconducting transport, whereas the LRS displays metallic behavior, consistent with conduction through vacancy-rich conductive pathways. Increasing $V_{Cl}$ concentration progressively weakens the temperature dependence of the HRS, which is consistent with the observed reduction of Schottky barrier height at the W/$KCuCl_{3-x}$ interface, decreasing from ~0.39 eV for HP1 to ~0.084 eV for HP5 (**Table S10 and Section 10, SI**).

The evolution of ionic transport was investigated by impedance spectroscopy (**Figure S11**). Increasing $V_{Cl}$ concentration results in a systematic decrease in HRS resistance accompanied by an increase in double-layer capacitance and a suppression of the Warburg impedance. These indicate enhanced charge accumulation and accelerated chloride-ion migration. Consistently, the calculated chloride-ion effective diffusion coefficient (**Table S11**) increases monotonically across the HP1-HP5 series, indicating progressively faster ionic transport as $V_{Cl}$ concentration increases.

Together, these complementary electrical measurements demonstrate that precursor-ratio-controlled vacancy engineering simultaneously lowers the transport barrier and accelerates chloride-ion migration, thereby governing the switching kinetics of $KCuCl_{3-x}$ memristors.

The influence of the enhanced current transport on the memristor programming characteristics was examined using pulse-width-dependent switching measurements (**Figure 6f**). The optimum programming pulse width decreases from 100 μs for HP1 to 50 μs for HP2, 10 μs for HP3, 5 μs for HP4, and 1 μs for HP5, indicating progressively faster switching kinetics as the $V_{Cl}$ concentration increases. The excellent agreement between the programming speed and the effective diffusion coefficients obtained from impedance spectroscopy provides evidence that precursor-ratio-controlled vacancy engineering effectively accelerates ion migration.

The influence of $V_{Cl}$ concentration on switching reliability was subsequently evaluated using pulse endurance and retention measurements performed under the optimum programming conditions for each composition (**Figures S12**). Endurance initially improves with increasing $V_{Cl}$ concentration and reaches a maximum for HP3, which maintains stable HRS and LRS states over more than $10^4$ switching cycles. However, further increasing the vacancy concentration progressively degrades cycling stability despite the faster switching speed. Similarly, although all devices maintain clearly distinguishable resistance states throughout the $10^4$ s retention period (**Figure 6i**), retention gradually deteriorates for HP4 and HP5.

These observations reveal a clear trade-off between switching kinetics and operational stability. While moderate vacancy concentrations provide sufficient mobile defects and controlled formation and rupture of localized conductive pathways, higher vacancy concentrations reduce switching efficiency and increase cycle-to-cycle variability.

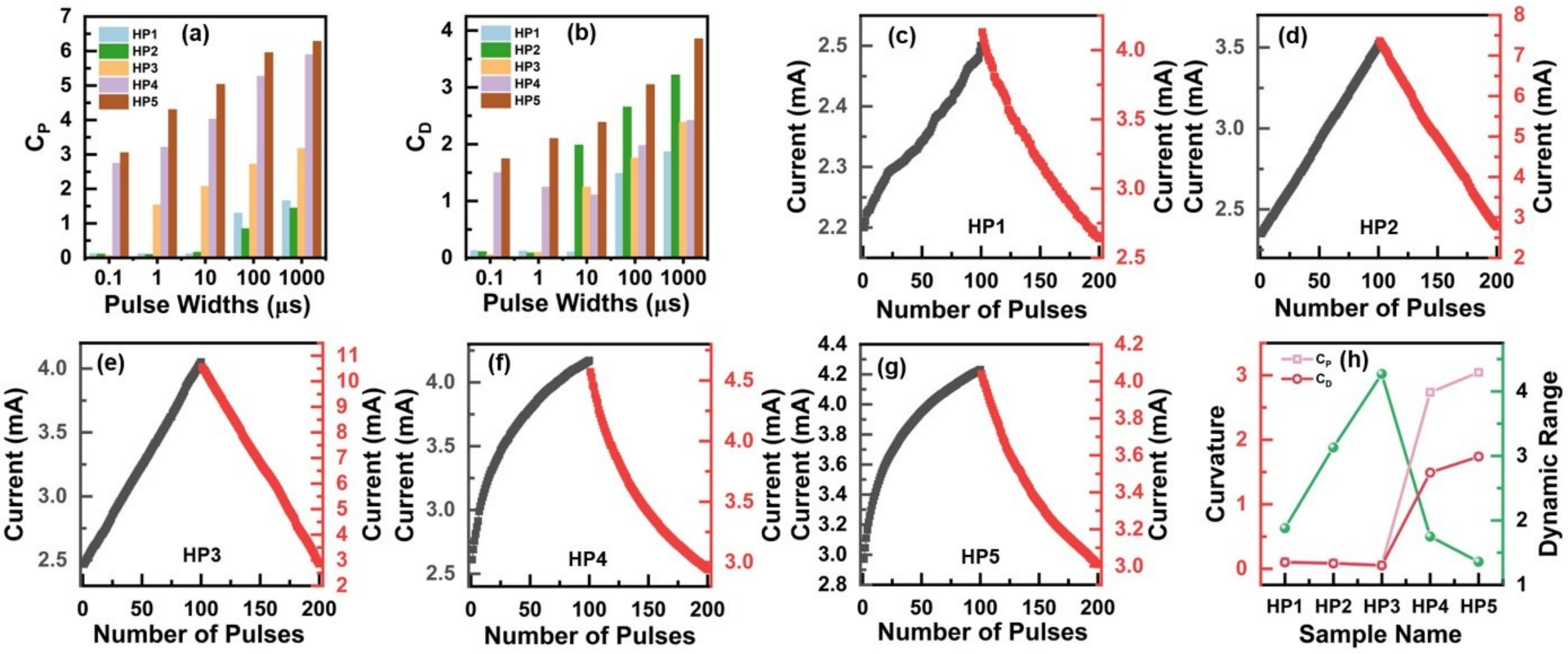

**Figure 7.** Effect of pulse width on **(a)** $C_P$ and **(b)** $C_D$. Gradual current modulation in W/$KCuCl_{3-x}$/FTO memristors under repeated pulse stimulation, demonstrating analog potentiation and depression behavior across **(c)** HP1, **(d)** HP2, **(e)** HP3, **(f)** HP4, and **(g)** HP5 devices. **(h)** Curvatures and dynamic range of P-D response at optimal pulse widths across HP1-HP5 devices

The influence of $V_{Cl}$ concentration on analog synaptic behavior was investigated using pulse voltage-driven potentiation (P) and depression (D) measurements. Based on the established switching characteristics, the programming pulse width was individually optimized for each composition while maintaining a constant programming amplitude of ±0.65 V (**Figures 7a-b**). The linearity of conductance modulation was quantified using the potentiation ($C_P$) and depression ($C_D$) curvature parameters, with values approaching zero corresponding to ideal analog weight updates (**Section S12**).

The optimum pulse width decreases gradually from 10 μs for HP1 to 1 μs for HP2 and further to 100 ns for HP3, which is consistent with the accelerated chloride-ion migration established from the transport measurements. However, HP4 and HP5 exhibit comparatively large $C_P$ and $C_D$ values even at the shortest available programming pulse width of 100 ns. This behaviour indicates that, beyond the optimum $V_{Cl}$ concentration, the higher vacancy density makes the conductance change induced by each programming pulse increasingly difficult to control, leading to rapid saturation of conductance and reduced linearity, as reported across various memristive materials systems.[51–53]

The optimized synaptic characteristics are summarized in **Figures 7c-h**. HP3 exhibits the lowest curvature parameters together with a relatively large dynamic range, demonstrating the most linear and symmetric conductance modulation among all compositions. In contrast, the increased curvature observed for HP4 and HP5 is consistent with their reduced switching stability and indicates increasingly nonlinear conductance evolution. These results demonstrate that an optimum $V_{Cl}$ concentration is essential for simultaneously maximizing programming speed and analog weight-update fidelity.

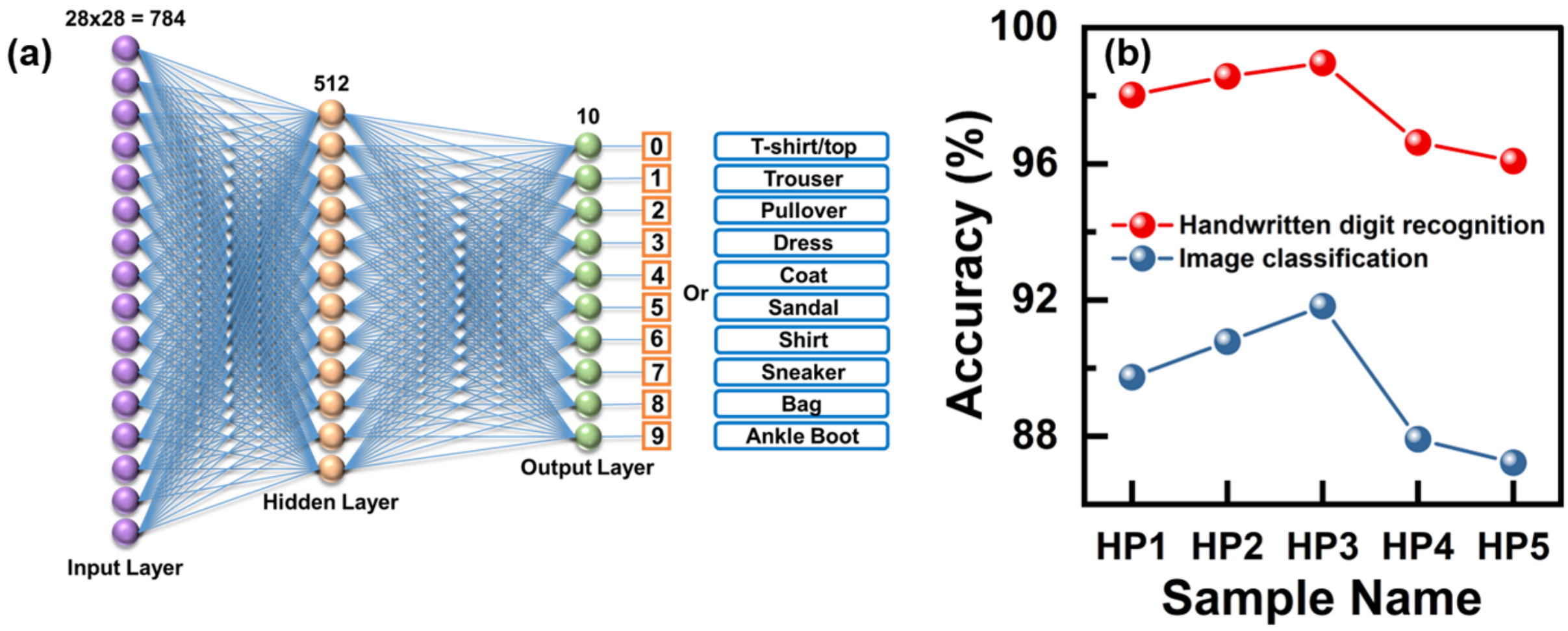

**Figure 8. (a)** Schematic representation of a three-layer artificial neural network (ANN) featuring an input layer, a hidden layer, and an output layer, and **(b)** the recognition and classification accuracies (%) of MNIST hand-written digits and MNIST fashion datasets for the HP1-HP5 devices.

To evaluate the practical implications of vacancy-engineered conductance modulation, the experimentally measured potentiation/depression characteristics were implemented as synaptic weights in artificial neural network simulations. The image recognition capabilities of the devices are assessed using a multilayer feed-forward neural network (NN) with one hidden layer, as illustrated in **Figure 8a**. The recognition accuracy obtained using the MNIST handwritten-digit dataset increases from 98.02% for HP1 to a maximum of 98.96% for HP3, which decreases to 96.63% and 96.08% for HP4 and HP5, respectively (**Figure 8b and S13**). A similar compositional dependence is observed for the more challenging Fashion-MNIST dataset, where the classification accuracy increases from 89.73% for HP1 to 91.82% for HP3, which decreases to 87.91% and 87.22% for HP4 and HP5.

The identical compositional trends observed across both datasets demonstrate that neuromorphic learning performance is primarily governed by the quality of analog conductance modulation. Near-linear potentiation and depression characteristics minimize weight-update errors and improve network convergence, whereas increasingly nonlinear conductance evolution at large vacancy concentrations reduces learning accuracy. The superior performance of HP3 is attributed to its optimal balance between transport kinetics, switching stability and analog conductance linearity.

The ANN simulation results, therefore, establish a direct structure-property-functionality relationship across the HP1-HP5 series. This demonstrates that optimized $V_{Cl}$ concentration with near-linear conduction modulation enables reliable translation of materials-level switching behavior into artificial intelligence (AI) integration potential.

## 3. Conclusion

In summary, we demonstrate that precursor-ratio-controlled non-stoichiometry provides a simple and effective strategy for systematically regulating $V_{Cl}$ concentration in lead-free $KCuCl_{3-x}$ thin films. Structural refinement reveals preferential vacancy formation at $Cl_T$ sites, while EPR and XPS analyses confirm vacancy-induced evolution of the $Cu^{2+}/Cu^{+}$ redox equilibrium. These defect-induced modifications are further reflected in the electronic structure through band-gap narrowing and enhanced electronic disorder. Among the investigated compositions, an intermediate $V_{Cl}$ concentration provides the most favorable balance between ionic mobility and switching controllability, resulting in forming-free gradual resistive switching, highly linear potentiation-depression characteristics and improved device reliability. When implemented as synaptic weights in artificial neural network simulations, the optimized memristor achieves recognition accuracies of approximately 98.96% and 91.82% on the MNIST and Fashion-MNIST datasets, respectively, demonstrating effective translation of device-level characteristics into learning performance. These findings establish anion-vacancy concentration engineering as a promising materials-design strategy for tailoring switching dynamics in lead-free halide memristors for neuromorphic applications.

## 4. Experimental Section

$KCuCl_{3-x}$ thin films were prepared by a scalable wet-chemical route and deposited on cleaned Fluorine-doped Tin Oxide (FTO, ~8 Ω $cm^{-2}$) substrates via spin-coating (WS-650-23 NPPB, Laurell Technologies), followed by low-temperature annealing. W/$KCuCl_{3-x}$/FTO memristor devices were fabricated by DC magnetron sputtering (Excel Instruments) of tungsten top electrodes through a shadow mask. To ensure environmental stability, the devices were encapsulated with PMMA as a protective layer. Detailed synthesis and device fabrication protocols are provided in **Section 1 of the Supporting Information (SI)**.

The structural and morphological properties, stoichiometry, and electronic bandgap of the thin films were characterized by X-ray diffraction (XRD) (PANalytical Empyrean), Rietveld Refinement (X'Pert Highscore Plus), X-ray Photoelectron spectroscopy (XPS) (PHI 5000 Versa Probe III, ULVAC PHI, USA), Scanning Electron Microscopy (SEM) (Zeiss Gemini), Atomic Force Microscopy (AFM) and Kelvin Probe Force Microscopy (KPFM) (Cypher S, Oxford Asylum), Continuous Wave Electron Paramagnetic Resonance spectroscopy (CW-EPR) (ELEXSYS 580, Bruker), and UV-Vis diffuse reflectance spectroscopy (UV-VIS-DRS) (Cary 5000, Agilent). KPFM measurements were performed using Ti/Ir-coated probes with a resonance frequency of 70.231 kHz and a spring constant of 2.07 N $m^{-1}$. For all samples, the scan rate, lift height, and applied tip bias were maintained at 1 Hz, 50 nm, and 3 V, respectively.

The neuromorphic functionalities of the W/$KCuCl_{3-x}$/FTO devices were investigated using electrical measurements with a Keithley 4200-SCS semiconductor parameter analyzer integrated with a JANIS probe station (ST500-UHT-1-(4CX), USA). The PMMA protective layer was readily punctured with tungsten probes to establish contact with the top electrodes, while the other regions remained protected from ambient moisture conditions, enabling reliable measurement of synaptic plasticity.

## 5. Supporting Information

Detailed synthesis, device fabrication protocols, Rietveld refinement, Phase purity, atomic% calculation from XPS, $V_{Cl}$ concentration calculation using XRD and XPS, $g_{eff}$ determination from EPR, AFM, cross-sectional SEM, Valence-band XPS and KPFM, Cumulative distribution plot of DC RS cycles, Electrode-area dependence and Temperature-dependence on LRS and HRS, Arrhenius plot, Schottky emission plot, and Impedance spectra of HRS studied across the HP1-HP5 series, Pulse endurance and retention characteristics of the fabricated memristors, Curvatures of P-D curves calculation, confusion matrices for each class in the datasets for all the devices, are provided in the Supporting Information.

## 6. Acknowledgment

SERB (CRG/2023/001265), SPARC (2024-2025/SEMI/P3099), Materials Science Centre, SEM facility under DST-FIST program, Central Research Facility, IIT Kharagpur, CMM acknowledges Prime Minister Research Fellowship (PMRF ID-2403448).